\documentclass[%
 aip,
 jmp,%
 amsmath,amssymb,
 reprint,%
]{revtex4-1}
\usepackage{graphics}
\usepackage{dcolumn}% Align table columns on decimal point
\usepackage{bm}% bold math
\usepackage{color}
\usepackage{graphicx}
\usepackage{soul}
\newcommand{\be}{\begin{equation}}
\newcommand{\ee}{\end{equation}}
\newcommand{\bea}{\begin{eqnarray}}
\newcommand{\eea}{\end{eqnarray}}
\newcommand{\beas}{\begin{eqnarray*}}
\newcommand{\eeas}{\end{eqnarray*}}

\newcommand{\nn}{\nonumber}

\begin{document}
\preprint{AIP/123-QED}
\title {Modeling the photoacoustic signal during  the  porous silicon formation}
\author{C. F. Ramirez-Gutierrez}
\affiliation{Posgrado en Ciencia e Ingenier\'ia de Materiales, Centro de F\'isica Aplicada y Tecnolog\'ia Avanzada, Universidad Nacional Aut\'onoma de M\'exico, Campus Juriquilla, C.P. 76230, Quer\'etaro,  Qro., M\'exico}
\author{J. D. Casta\~no-Yepes}
\affiliation{Instituto de Ciencias Nucleares, Universidad Nacional Aut\'onoma de M\'exico,  M\'exico Distrito Federal, C. P. 04510, M\'exico}
\author{M. E. Rodriguez-Garc\'ia}
\altaffiliation[E-mail corresponding author: ]{marioga@fata.unam.mx}
\affiliation{Departamento de Nanotecnolog\'ia, Centro de F\'isica Aplicada y Tecnolog\'ia Avanzada, Universidad Nacional Aut\'onoma de M\'exico,  Campus Juriquilla, C.P. 76230, Qro., M\'exico}
%%%%%%%%%%%%%%%%%%%%%%%%%%%%%%%%%%%%%%%%%%%%%%%%%%%%%%%%%%%%%%%%%%%%%%%%%%%%%%%%%%%%%%%%%%%%%%%
\begin{abstract}
Within this work, the kinetics of the growing stage of porous silicon (PS) during the etching process was studied using the photoacoustic technique. A p-type Si with low resistivity was used as a substrate. An extension of Rosencwaig and Gersho model is proposed in order to analyze the temporary changes that take place in the amplitude of the photoacoustic signal during the PS growth. The solution of the heat equation takes into account the modulated laser beam, the changes in the reflectance of the PS-backing heterostructure, the electrochemical reaction, and the Joule effect as thermal sources. The model includes the time-dependence of the sample thickness during the electrochemical etching of PS. The changes in the reflectance are identified as the laser reflections in the internal layers of the system. The reflectance is modeled by an additional sinusoidal-monochromatic light source  and its modulated frequency is related to the velocity of the PS growth. The chemical reaction and the DC components of the heat sources are taken as an average value from the experimental data. The theoretical results are in agreement with the experimental data and  hence provided a method to determine variables of the PS growth, such as the etching velocity and the thickness of the porous layer during the growing process.
\end{abstract}

\maketitle
%%%%%%%%%%%%%%%%%%%%%%%%%%%%%%%%%%%%%%%%%%%%%%%%%%%%%%%%%%%%%%%%%%%%%%%%%%%%%%%%%
\section{Introduction}
Today, the increase of the demand of electronic devices based on porous silicon (PS) for biological, medical, and sensor applications, is an important topic. This means that it is necessary to develop techniques that allow the \textit{in situ} study of the PS formation during the electrochemical etching.\cite{Uhlir} The intrinsic and extrinsic parameters affect the physical properties of PS.  The external parameters, such as anodization current, electrolyte composition, temperature, and  intrinsic parameters, such as  resistivity, crystalline orientation, carrier uniformity, and defects.\cite{Foll,Zhang} The main problem for the fabrication of devices based on PS is its reproducibility, and usually, the  characterization of its properties is performed \textit{ex situ}. There are a few works about the \textit{in situ} techniques to study the PS formation.\cite{Zhao,RAO,ISA} Gaburro et al.\cite{Gaburro} used a method based on optical interferometry using front detection. They study the changes in the optical path during the etching process. It was found that the etching velocity as well as the porosity have time dependence. On the other hand, Foss et al. \cite{Foss} used the same methodology  to determine the porosity, the thickness, and the roughness of the PS film  thru a back detection.

However,  in the referred works, the information regarding the growing process is neglected. Ramirez-Gutierrez et al.,\cite{RCR} reported a methodology based on the photoacoustic (PA) technique for monitoring \textit{in situ} the PS etching. PA has been used to study kinetic processes, thermal and electronic properties of semiconductor materials.\cite{REVMario, Calderon, Gutierrez,Espinosa}Therefore, it is expected that the  PA technique could be able to detect \textit{in situ} the chemical reaction and the changes in the optical properties during the PS formation.\\

In the previous works, the PA signal during the electrochemical etching of PS formation had not been modeled because the thermal sources that influence the changes in the PA signal were unknown. The primary mechanism to generate the PA signal is the heat transfer between the different regions of the cell; therefore, if there is control over each one of the parameters that affect the PS growing (solvent/acid ratio, temperature, current, among others) as well as a complete knowledge of the Si substrate\cite{JAPMario} (carrier uniformity), it is possible to model the PA signal by solving the heat equation. The first thermal  source for the PA signal are the modulated laser, the second  thermal source is the changes in the reflectance due to there is a kinetic process in which the thickness of  the PS sample changes as a function of time, bringing about variations in the reflectance of  the incident modulated beam that behaves as an interference transmission filter.\cite{RCR} It is clear that the chemical reaction that allows the PS formation is also a thermal source as well as the Joule effect. 
The heat originated from the electrochemical reaction  depends on the carrier uniformity of the sample.\cite{JAPMario} These sources have to be taken into account for a theoretical description of the PA phenomena. Nevertheless, Dramicanic et al.\cite{Dramicanic} described  a theory for the PA signal in semicondcutors (Ge) to determine its thermal and electronic properties,  but they assumed that the sample is uniformly dopped.\cite{Dramicanic}

In this work, a phenomenological extension of the Rosencwaig and Gersho (RG) model\cite{Gersho} is used to model the PA amplitude signal during the PS formation. The heat sources are explicitly written and specified by the incidence, the reflection, and the absorption of light by the interfaces formed during the etching process. The heat from the electrochemical reaction, the Joule effect (DC parameters), and the velocity of PS formation are taken as fitting parameters. The paper is organized as follows: In Sec. \ref{SectionI} the extension of the RG photoacoustic model is developed with the introduction of thermal sources. The PA amplitude is modeled by the solution of the heat equation in one dimension for the three layers system. In Sec. \ref{experimentalsection} the experimental details of the implementation of the PA technique are presented. It shows that the total contribution of the thermal inertia, the Joule effect and the  etching process as a function of time can be obtained from the scaling of the PA signal baseline. In Sec. \ref{results} the results of the PA model are discussed and compared with the experimental data finding a good agreement between them. Finally, in Sec. \ref{conclusions}, the summary and the conclusions are presented.

%%%%%%%%%%%%%%%%%%%%%%%%%%%%%%%%%%%%%%%%%%%%%%%%%%%%%%%%%%%%%%%%%%%%%%%%%%%%%%%%%
\section{Photoacoustic Model} \label{SectionI}

\begin{figure}
{\centering
\includegraphics[scale=0.3]{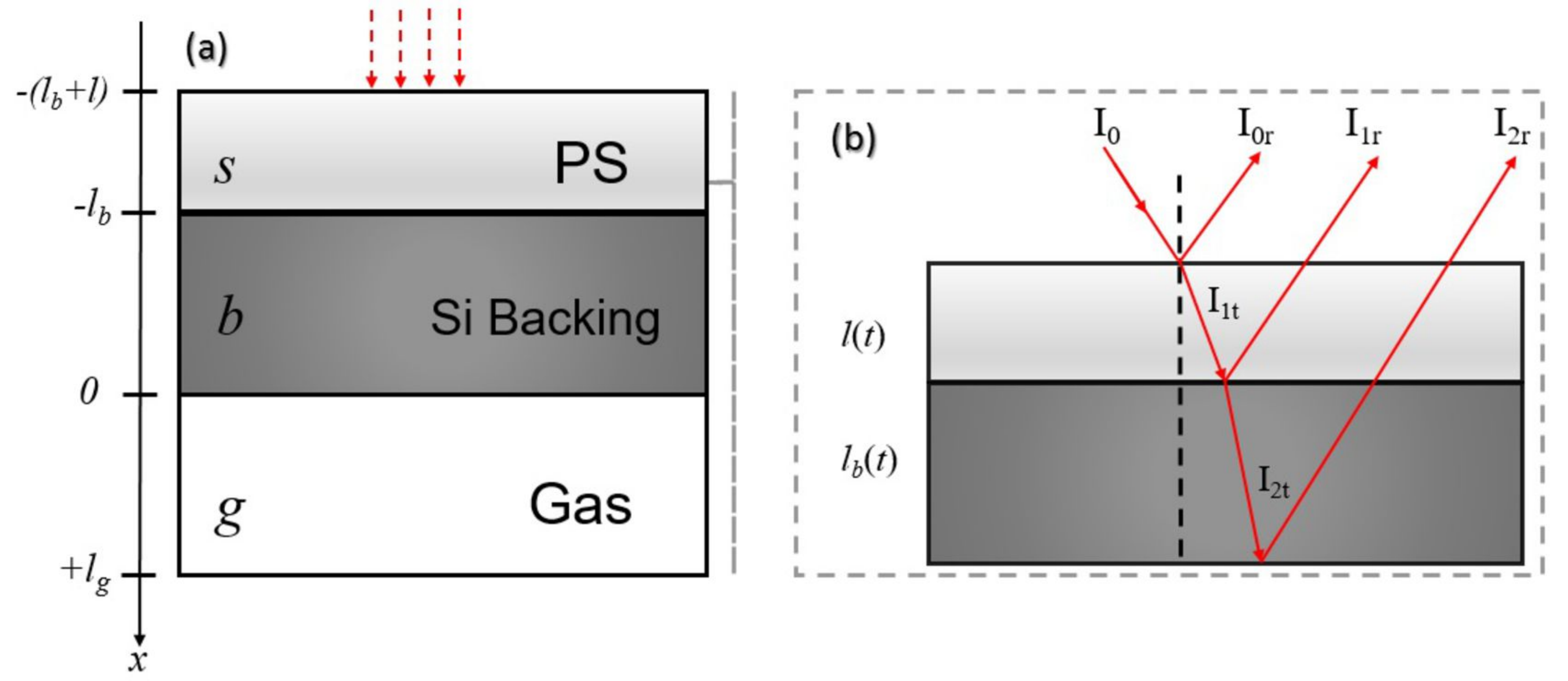}
\caption{Cross-section of the photoacoustic cell. The labels $s$, $b$, and $g$ correspond to sample, backing, and gas, respectively.  
\label{cell}}}
\end{figure}

A detailed description of the photoacoustic setup was reported elsewhere.\cite{RCR} In this experiment the incident radiation is from the front, and the photoacoustic detection is located at the back of the Si sample as it is seen in Fig.~\ref{cell}(a). It is also considered a cylindrical photoacoustic cell with a  cross-section geometry as it is shown in Fig.~\ref{cell}. It is assumed  that $l$, $l_b$, and $l_g$ are the thickness of the sample (s), the backing material (b), and the  gas column (g), respectively. The heat transfer is considered one-dimensional in the $x$-direction,\cite{Dramicanic} and the dilatation/contraction by the thermal effects is ignored.
%%%%%%%%%%%%%%%%%%%%%%%%%%%%%%%%%%%%%%%%%%%%%%%%%%%%%%%%%%%%%%%%%%%%%%%%%%%%%%%%%
%%%%%%%%%%%%%%%%%%%%%%%%%%%%%%%%%%%%%%%%%%%%%%%%%%%%%%%%%%%%%%%%%%%%%%%%%%%%%%%%%

The electrochemical reaction causes the sample-backing boundary $l_b$ to move in time. It is considered that $v$ is the velocity of the PS formation which can be taken as a constant~\cite{RCR}. Thus, the moving boundary can be written as:\\
\bea
l_b(t)=l_b-vt,
\label{MovingBoundary}
\eea
giving as a result two scenarios: the first one is when the reaction has not consumed the backing and the second is when there is only PS. It implies that there is an interval of time $T=l_b/v$ where there are three regions (sample, backing, and gas) and for times $t>T$ there are only two parts (sample and gas).
The thermal sources arise from the periodic excitation of the sample and the electrochemical etching in the sample-backing interface. The sample heats the backing material which in turn modifies the temperature distribution of the surrounding gas where the microphone is fixed for the signal detection. Thus, the heat equation for the sample is:
\bea
\nabla^2T_s(x,t)-\frac{1}{\alpha_s}\frac{\partial T_s(x,t)}{\partial t}=f_s(x,t)+Q(t,x),
\label{ecs}
\eea
where $f_s(x,t)$ and $Q(t,x)$  are the heat sources  identified as the modulated laser and the  electrochemical reaction respectively. The electrochemical reaction occurs only at the PS/Si interface. 

In order to determine $f_s(x,t)$, it is assumed that a modulated light source impings the surface at a frequency $\omega_1$ with an intensity of $I_0$.The incident radiation ($I_0$) is in part reflected ($I_{0r}$) and transmitted ($I_{1t}$). At the interface sample-backing, the transmitted light is reflected ($I_{1r}$) to the sample and transmitted ($I_{2t}$) to the backing (Fig.~\ref{cell}(b)). The backing thickness changes as a function of time by the electrochemical etch. The reflected light appears with  a different phase in relation to the incident light to satisfy the boundary conditions of the electromagnetic field. This effect can be represented by an additional light source with frequency $\omega_2$ in the backing-sample direction. The superposition of these two effects allows to write $f_s(x,t)$ in the region $-(l+l_b)<x<-l_b+vt$ as:
\bea
f_s(x,t)=&-&A(1+e^{i\omega_1t})e^{-\beta_s (x+l_b+l)}\nn\\
&-&B(1+e^{i\omega_2t})e^{\beta_s (x+l_b+vt)},
\label{sources}
\eea
where:
\bea
A=\frac{\beta_s I_{1t} \eta_s}{2k_s} \;\;\text{and}\;\; B=\frac{\beta_s I_{1r} \eta_s}{2k_s}
\label{coeff1}
\eea
represent the heat density at $x=-(l+l_b)$ and $x=-l_b$, respectively, with $\eta_i$ the efficiency in which the absorbed light is converted to heat by the nonradiative deexcitation processes.\cite{Gersho} Also, $\alpha_i$ represents the total thermal diffusivity, $k_i$ is the total thermal conductivity,  and $\beta_i$  is the optical absorption coefficient for each material.

For the backing material in the region $-l_b+vt<x<0$:

\bea
\nabla^2T_b(x,t)-\frac{1}{\alpha_b}\frac{\partial T_b(x,t)}{\partial t}&=&f_b(x,t)
\label{ecb}
\eea
and its heat source term:
\bea
f_b(x,t)=&-&C(1+e^{i\omega_1t})e^{-\beta_b (x+l_b+vt)}\nn\\
&-&D(1+e^{i\omega_2t})e^{\beta_bx},
\label{sourceb}
\eea
where $C$ and $D$ have the same form that Eq.~(\ref{coeff1}) with the respective index and intensities. The $\beta_b$~\cite{Martin}is $8.12\times10^{2}$ cm$^{-1}$ and the thermal conductivity can take values between 1 to 1.3 W$/$K cm in this simulation ( see Table~\ref{tabla1}).  These intensities satisfy the system of equations given by:

\bea
I_{1t}+I_{r0}&=&I_{0}\nn\\
I_{2t}+I_{1r}&=&I_{1t}\nn\\
I_{2t}+I_{2r}&=&1,
\label{intensidades}
\eea
where $I_{0}$ and $I_{r0}$ can be extracted from the experiment. Finally, the heat equation for the gas is given by:
\bea
\nabla^2T_g(x,t)-\frac{1}{\alpha_g}\frac{\partial T_g(x,t)}{\partial t}=0.
\label{ecg}
\eea
To solve the Eqs. (\ref{ecs}), (\ref{ecb}), and (\ref{ecg}) we use the superposition principle for each source of heat by an extension of the (RG) model.\cite{Gersho} The temperature field in the cell is given by:
\bea
T(x,t)=\text{Re}\left\{\phi(x,t)\right\}+T_0,
\label{Tgeneral}
\eea
where $T_0$ is the room temperature, this allows us to write the solution of $\phi_i$ for each region as:
\bea
\phi_g(x,t)&=&
\tilde{\phi}_g(x,t)+\sum_{n=1}^2\theta_n\exp\left[-\sigma^n_gx+i\omega_nt\right],\nn\\
\label{phig}
\eea
\bea
\phi_b(x,t)&=&%
\tilde{\phi}_b(x,t)\nn\\
&+& \left[U_1e^{-\sigma_b^1(v)(x+l_b+vt)}-V_1e^{-\beta_b (x+l_b+vt)}\right]e^{i\omega_1t}\nn\\
&+&\left[U_2e^{\sigma_b^2(0)x}-V_2e^{\beta_bx}\right]e^{i\omega_2t},
\label{phib}
\eea
and
\bea
\phi_s(x,t)&=&
\phi^{\text{DC}}_s(x,t)-\alpha_s                              \int_{0}^t Q(\tau)d\tau\nn\\
&+& \left[W_1e^{-\sigma_s^1(0)(x+l+l_b)}+W_2e^{\sigma_s^1(0)(x+l+l_b)}\right.\nn\\
&-&\left.E_1e^{-\beta_s (x+l+l_b)}\right]e^{i\omega_1t}\nn\\
&+&\left[W_3e^{-\sigma_s^2(v)(x+l_b+vt)}+W_4e^{\sigma_s^2(v)(x+l_b+vt)}\right.\nn\\
&-&\left.E_2e^{\beta_s (x+l_b+vt)}\right]e^{i\omega_2t},
\label{phis}
\eea
in which we have defined:
\bea
\tilde{\phi}_i(x,t)&=&\text{DC component of the PA signal}\nn\\&+&\text{Etch heating contribution}\nn\\&+&\text{Joule effect from the current source}\nn\\
&\equiv&\Delta P_{\text{DC}}+\text{Etch},
\label{phitilde}
\eea
where the contributions from the etching heat and the the Joule effect have been included with the label Etch. The coefficients $W_n$ and $U_n$ are obtained from the temperature and the flux continuity for each AC component of Eqs.~(\ref{phig})-(\ref{phis}). It is easy to show that $\sigma$'s are related to the thermal diffusivity $\alpha$ as follows:
\bea
\sigma_g^n&=&(1+i)\sqrt{\frac{\omega_n}{2\alpha_g}}\nn\\
\sigma_j^n(v)&=&\frac{1}{2}\left(\frac{v}{2}+\frac{\sqrt{v^2+4i\alpha_j\omega_n}}{\alpha_j}\right), \;\; j=b,s.
\label{sigmas}
\eea
The constants $V_n$ and $E_n$ are determined by using Eqs.~(\ref{phib}) and~(\ref{phis}) in Eqs.~(\ref{ecb}) and~(\ref{ecs}), which give us:
\bea
V_1=\frac{C}{\beta_b^2+\frac{\beta_b v-i\omega_1}{\alpha_b}},\;\;V_2=\frac{D}{\beta_b^2-\frac{i\omega_2}{\alpha_b}}
\label{V1V2}
\eea
and
\bea
E_1=\frac{A}{\beta_s^2-\frac{i\omega_1}{\alpha_s}},\;\;E_2=\frac{B}{\beta_s^2+\frac{\beta_s v-i\omega_2}{\alpha_s}}
\label{E1E2}
\eea
The temperature and flux continuity conditions at the sample surfaces are explicitly given by:
\bea
\phi_g(0,t)&=&\phi_b(0,t)\nn\\
\phi_b(-l_b+vt,t)&=&\phi_s(-l_b+vt,t)\nn\\
k_g\frac{\partial\phi_g}{\partial x}(0,t)&=&k_b\frac{\partial\phi_b}{\partial x}(0,t)\nn\\
k_b\frac{\partial\phi_b}{\partial x}(-l_b+vt,t)&=&k_s\frac{\partial\phi_s}{\partial x}(-l_b+vt,t),
\label{conditions}
\eea
which we apply separately to each source and to their  AC and DC components. Our attention is focused on $\theta_n$, the complex amplitude of the periodic temperature at $x=0$, which give us the source of photoacoustic signal. By using the Eqs.~(\ref{conditions}), these amplitudes are:
\bea
\theta_1&=&\frac{k_b(\beta_b-\sigma_b^1(v))}{k_b\sigma_b^1(v)-k_g\sigma_g^1}V_1e^{-\beta_b(l_b-vt)},\nn\\
\theta_2&=&\frac{k_b(\beta_b-\sigma_b^2(0))}{k_b\sigma_b^2(0)+k_g\sigma_g^2}V_2.
\label{Thetas}
\eea
The periodic excitation of the backing-gas boundary expands the gas column  and contract at the same frequency of the oscillations of the temperature field. The PA amplitude signal is time dependent; therefore, it is convenient to define the average temperature of the gas. From the functional form of the temperature field, there is a characteristic length $\lambda_g$ in which the periodic temperature variation in the gas is completely damped out. As a result, the gas average temperature is defined as:
\bea
\bar{\phi}(t)\equiv\frac{1}{\lambda_g}\int_0^{\lambda_g}dx\;\phi_{AC}(x,t),
\label{averagetemperature}
\eea
from Eq.~(\ref{phig}) this gives:
\bea
\bar{\phi}(t)&=&\sum_{n=1}^2\frac{\theta_n}{2}\sqrt{\frac{2\alpha_g}{\omega_n}}(1-i)\left(1-e^{-\sigma_g^n\lambda_g}\right)e^{i\omega_n t}\nn\\
&\approx&\sum_{n=1}^2\sqrt{\frac{\alpha_g}{\omega_n}}\theta_ne^{i(\omega_n t-\pi/4)}
\label{averageg}.
\eea
By using the ideal gas law, the displacement $\delta x(t)$ of a column of gas is given by:

\bea
\delta x(t)=\lambda_g \sum_{n=1}^2\sqrt{\frac{\alpha_g}{\omega_n}}\frac{\theta_n}{\Phi}e^{i(\omega_n t-\pi/4)},
\label{desplazamiento}
\eea
where we have set the average DC temperature of the gas boundary layer equal to the DC temperature at the solid surface:
\bea
\Phi=\theta_0+T_0.
\label{DC-Component}
\eea

The pressure at time $t$ can be derived if it is  assumed that the gas is compressed adiabatically. From the adiabatic gas law and along with Eq.~(\ref{desplazamiento})   is obtained:

\bea
\delta P(t)&=&\gamma\frac{P_0}{V_0}\delta V(t)\nn\\
&=&\gamma P_0\frac{\lambda_g}{l_g}\sum_{n=1}^2\sqrt{\frac{\alpha_g}{\omega_n}}\frac{\theta_n}{\Phi}e^{i(\omega_n t-\pi/4)},
\label{incrementalpressure}
\eea
where $\gamma$ is the adiabatic index. Finally, $\Delta P(t)$ is modeled by using the Eq.~ref{Rep} and detected by the microphone at $x_{\text{mic}}$ and converted in the photoacoustic signal.
\bea
\Delta P(t)= \Re\{\delta P(t)\}.
\label{ReP}
\eea
%%%%%%%%%%%%%%%%%%%%%%%%%%%%%%%%%%%%%%%%%%%%%%%%%%%%%%%%%%%%%%%%%%%%%%%%%%%%%%%%%
\section{Experimental Details}\label{experimentalsection}

\begin{figure}[h]
{\centering
\includegraphics[scale=0.32]{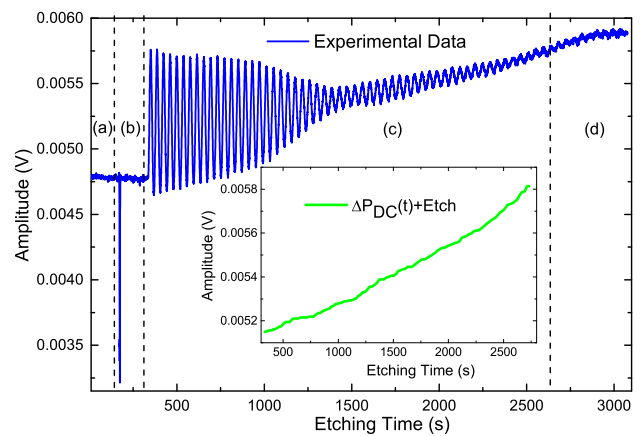}
\caption{(Color line) The experimental data of the PA signal as a function of the etching time (blue). The inset shows the contribution from the DC component extracted from the experimental data (green).} \label{experimental}}
\end{figure}

The description of the heating process required knowing the contribution to the heat transfer of the chemical etching and the thermal inertia, or the DC component, which can be obtained from the experimental PA amplitude signal showed in Fig.~\ref{experimental}. This figure shows a characteristic PA Amplitude signal for the PS formation using a P-type Si wafer, $0.005$ $\Omega$cm (WRS Materials, USA) as a substrate, in which the following parameters were used: 500 $\mu$m thickness, $\left[100\right]$ orientation. The spot size of the focused laser beam was $1000$ $\mu$m  to satisfy  the one dimensional  model, which means that the electronic carrier diffusion $(\mu)$ and the thermal diffusion length $(\mu_{th})$ are less than the beam spot size.\cite{Dramicanic} It is important to remember that  $\mu=\sqrt{D_{n,p}\tau}$,  where $D_{n,p}$ is the electronic carrier diffusion coefficient,  $\tau$ is the carrier life time,  and $\mu_{th}=\sqrt{\alpha_i/\pi f}$.

The laser wavelength used  in this experiment was $808$ nm which is super-band gap for Si (1.12 eV) and sub band gap for PS that is typically about 2.2 eV -3.2 eV depending on its porosity.\cite{Chen} The etching conditions were 7:3 V/V ethanol/HF ratio, and 20 mA/cm$^2$ current density.\\

The PA amplitude signal was divided into four regions (See Fig.~\ref{experimental}). Region (a) in which the Si sample is directly impinged by the laser, the part (b) is characterized by a drastic change in the PA signal which is related to the emptying of the electrolyte into the chamber. A pause takes place while the native surface of silicon oxide reacts with the HF and it is removed from the surface.  At this stage, the power supply is still off. In the region (c) the power supply is turned on and the PS formation takes place. This region is characterized by the oscillatory shape of the PA signal. Finally, for region (d) the power supply is still turned on, but the coherent interference effect is lost due to the thick film\cite{Siapkas} of PS. The inset in Fig.~\ref{experimental} is the total contribution to the PA signal from the non-radiative process as a function of the etching time. The way in which this curve was obtained is explained in Sec. IV. It is suggested  the Ref. \cite{RCR} for  a more detailed  description of the electrochemical PA experimental setup.

%%%%%%%%%%%%%%%%%%%%%%%%%%%%%%%%%%%%%%%%%%%%%%%%%%%%%%%%%%%%%%%%%%%
 %%%%%%%%%%%%%%%%%%% Tabla %%%%%%%%%%%%%%%%%%%

\begin{table*}
\caption{\label{tab:table1} Reported values for the thermal and electric properties of  Si  at different doped levels.}
\label{tabla1}
\begin{ruledtabular}
\begin{tabular}{ccccccccc}
Type&Dopant concentration&Resistivity (300K)&Thermal conductivity&Thermal diffusivity&$D_{n,p}$&$\tau$&Reference\\
 
 \,&cm$^{-3}$& $\Omega$ cm&W K$^{-1}$ cm$^{-1}$&cm$^{2}$ s$^{-1}$&cm$^2$s$^{-1}$&$\mu$s
 \\ \hline

n&$10^{15}$&0.025&-&$0.795$&-&-&Amato et. al. \cite{Amato}\\
n&$5\times10^{5}$&1&-&0.797 &-&-&Amato et. al. \cite{Amato}\\
n&$1.5\times10^{14}$&25&-&0.821&-&-&Amato et. al. \cite{Amato}\\
n&$5.7\times10^{13}$&75&-&0.868&-&-&Amato et. al. \cite{Amato}\\
n&$10^{12}$&4000&-&0.985&-&-&Amato et. al. \cite{Amato}\\
p&$-$&14-24&-&0.80&30&10.5&Salnick et. al. \cite{Salnick}\\
p&$-$&14-24&-&0.75&5.5&95&Rodriguez et. al.\cite{Mariovalexp2}\\
p&$-$&25-44&-&0.96&3.1&1400&Rodriguez et. al.\cite{Mariovalexp2}\\
p&$1.5\times 10^{14}$&88.9&-&0.76&5&100&Pinto et. al.\cite{Pinto}\\
p&$1\times 10^{16}$&0.52&1.48&-&-&-&Burzo et. al.\cite{Burzo}\\
p&$2\times 10^{19}$&0.005&1.22&-&-&-& Burzo et. al. \cite{Burzo}\\

\end{tabular}
\end{ruledtabular}
\end{table*}
%%%%%%%%%%%%%%%%%%%%%%%%%%%%%%%%%%%%%%%%%%%%%%%%%%%%%%%%%%%%%%%%%%%%

\begin{table}[h]
\caption{Values for the relevant parameters used in Eq.~(\ref{ReP}) in order to reproduce Fig.~\ref{PAAC}.}
\label{tabla2}
\centering
\begin{tabular}{p{2cm} p{2.3cm} p{2cm}}

\hline
\hline 
$C$                  &$2\times 10^6$        &K/m$^2$ \\
$D$                  &$2\times 10^6$        &K/m$^2$ \\
$\omega_1$           & $26\;\pi$            &rad/s \\
$\omega_2$           & $5\times 10^{-2}\pi$ &rad/s\\
$\alpha_g$\cite{Hand}&0.20                  &cm$^2$/s\\
$\kappa_g$\cite{Hand}&$2.53\times 10^{-4}$  &W/(cmK)\\
$\beta_b(808\,\text{nm})$&$8.12\times10^2$  &cm$^{-1}$\\
$\beta_s(808\,\text{nm})$&$2.20\times10^2$  &cm$^{-1}$\\
$l_b$                &$500\times 10^{-6}$   &m \\
$x_{\text{mic}}$     &$2\times 10^{-6}$     &m \\
$v$                  & $8$                  &nm/s \\

\hline \hline
\end{tabular}
\end{table}
%%%%%%%%%%%%%%%%%%%%%%%%%%%%%%%%%%%%%%%%%%%%%%%%%%%%%%%%%%%%%%%%%%%%%
\section{Results and discussion}\label{results}
The modified RG model for the PA signal takes into account the above-mentioned thermal sources. By using experimental values of the  PA cell size, the modulated frequency, the thermal properties of the sample and gas, and  the  sample size, it is possible to calculate the AC component of the PA amplitude signal during PS growing (Eq.~\ref{DC-Component}). Table~\ref{tabla1} shows the experimentally reported values for the electronic and thermal parameters for p and n Si. The velocity of the chemical reaction is involved in the PA amplitude signal, 
such is analyzed in detail in the next section.\\
%%%%%%%%%%%%%%%%%%%%%%%%%%%%%%%%%%%%%%%%%%%%%%%%%%%%%%%%%%%%%%%%%%%%
\begin{figure}[h]
{\centering
\includegraphics[scale=0.43]{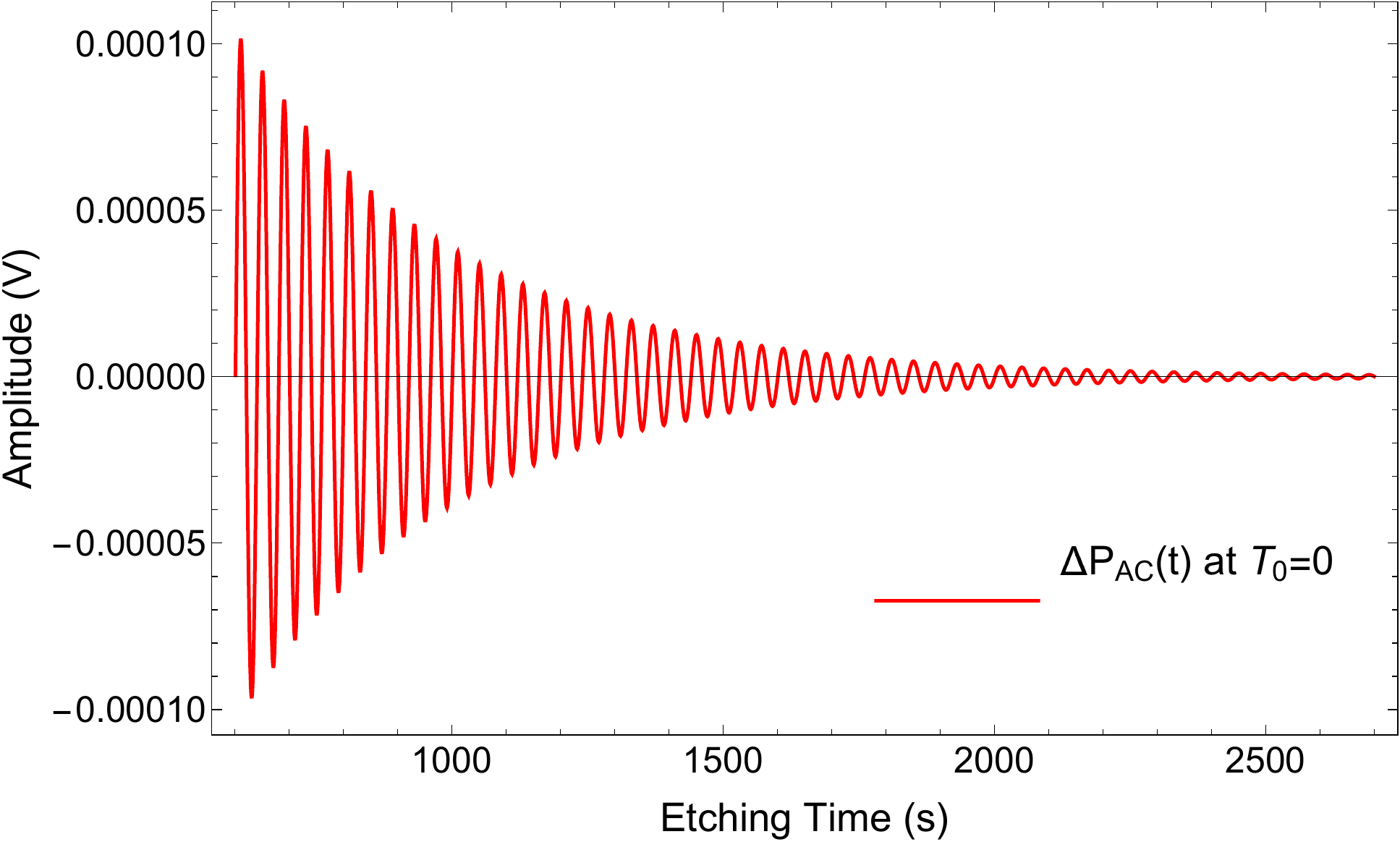}
\caption {AC-photoacoustic signal computed from Eq.~(\ref{ReP}), at $T_0=0$, in the interval of etching $450<t<2700$. It is observed that the modulation comes from the frequency $\omega_2$, i. e., the effect of reflectance in the PS which is related to a period $T\sim 40$s.\label{PAAC}}}
\end{figure}

%%%%%%%%%%%%%%%%%%%%%%%%%%%%%%%%%%%%%%%%%%%%%%%%%%%%%%%%%%%%%%%%%%%%
Fig.~\ref{PAAC} shows the AC component of the PA amplitude signal calculated from Eq.~(\ref{ReP}) as a function of the etching time at $T_0=0$. This simulation  includes the values shown in Tables~\ref{tabla1} and~\ref{tabla2}. The frequency of this signal is constant in time. That means that one of the real parts of the amplitudes in Eq.~(\ref{Thetas}) predominates in the heat transfer. Also, the amplitude of this signal is modulated and it is  related to the increase in the optical path of the transmitted light mainly in the PS layer. The baseline in Fig.~\ref{PAAC} is centered at zero, given that the modulated source does not produce an average heating of the sample since there are continuous excitation/deexcitation processes.
%%%%%%%%%%%%%%%%%%%%%%%%%%%%%%%%%%%%%%%%%%%%%%%%%%%%%%%%%%%%%%%%%%%%
\begin{figure}[h]
{\centering
\includegraphics[scale=0.43]{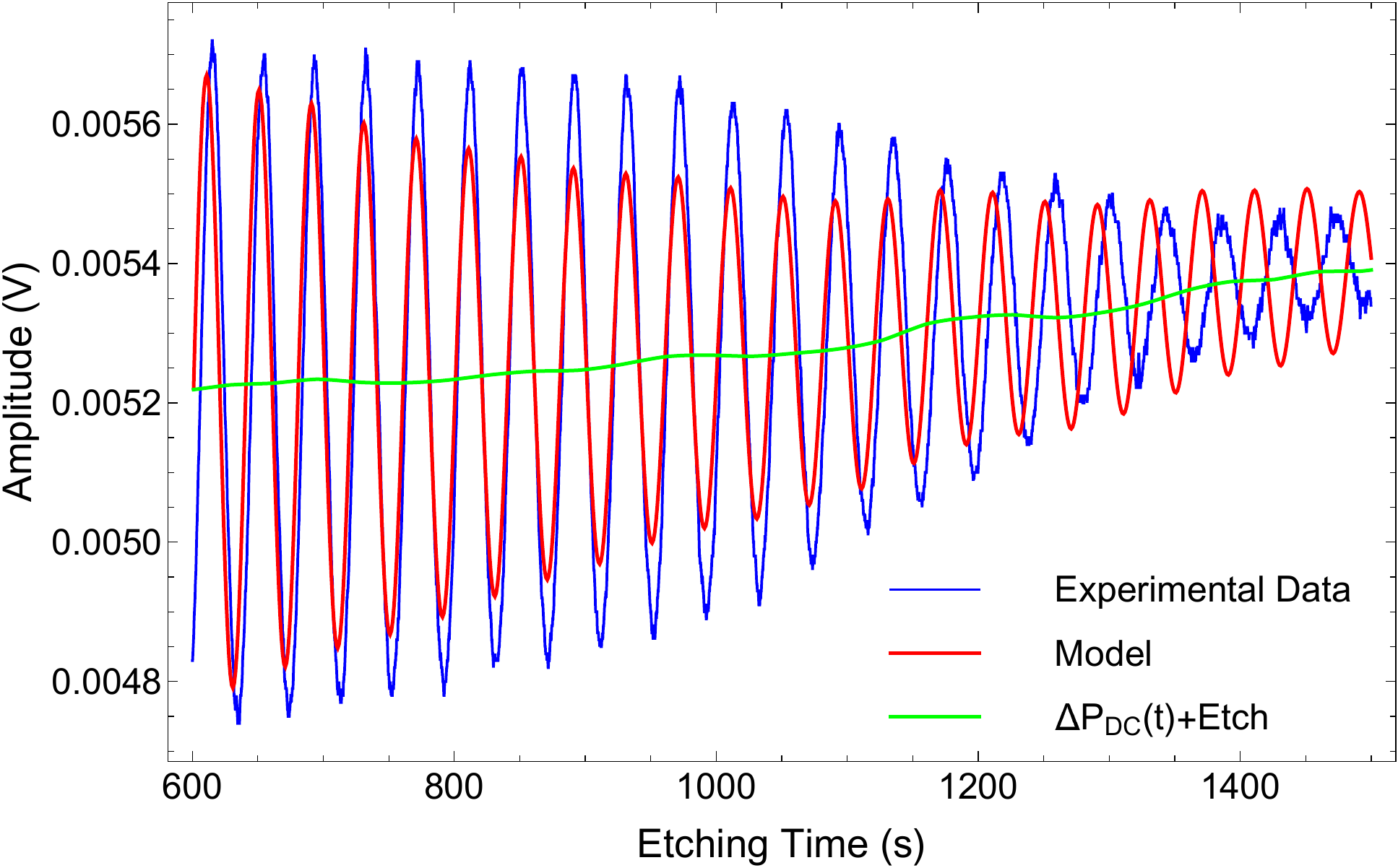}
\caption{(Color line) Photoacoustic signal of experimental data (blue) and phenomenological model (red). Also the contribution of etch and DC component is shown (green) for the etching time $450\;\text{s}<t<1500\;\text{s}$.\label{P1}}}
\end{figure}
%%%%%%%%%%%%%%%%%%%%%%%%%%%%%%%%%%%%%%%%%%%%%%%%%%%%%%%%%%%%%%%%%%%%
\begin{figure}[h]
{\centering
\includegraphics[scale=0.43]{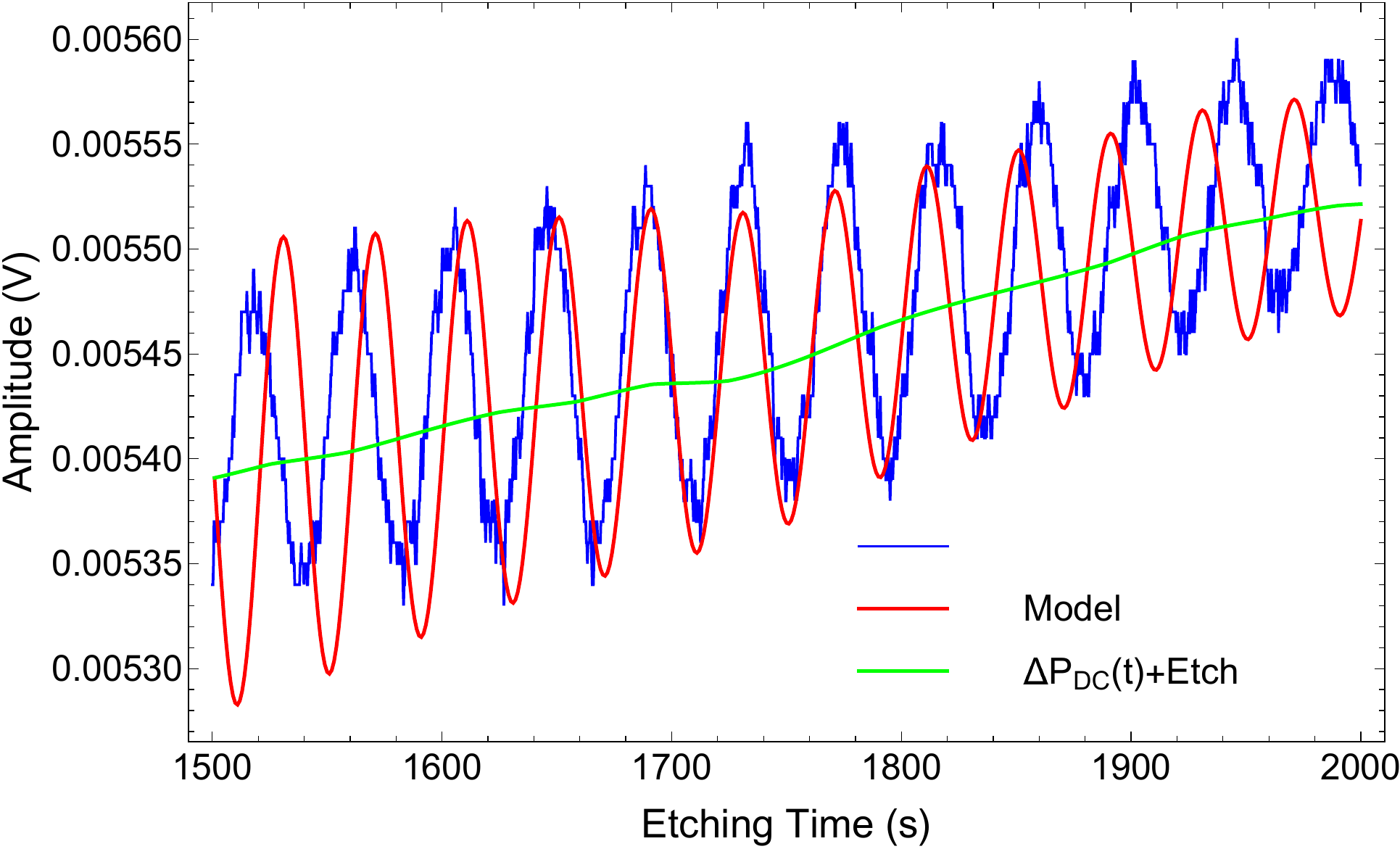}
\caption{(Color line) Photoacoustic signal of experimental data (blue) and phenomenological model (red). Also the contribution of etch and DC component is shown (green) for the etching time $1500\;\text{s}<t<2000\;\text{s}$.\label{P2}}}
\end{figure}
%%%%%%%%%%%%%%%%%%%%%%%%%%%%%%%%%%%%%%%%%%%%%%%%%%%%%%%%%%%%%%%%%%%%

Comparing  Figs.~\ref{experimental} and \ref{PAAC}, it can be seen that the predominant amplitude in this signal comes from $\theta_2$  (Eq. 20) given that the AC-PA cycles correspond to the frequency $\omega_2$ in accordance with the experimental data in which a complete  cycle approximates $t_{\text{etching}}\sim 40$ s.

\begin{figure}[h]
{\centering
\includegraphics[scale=0.43]{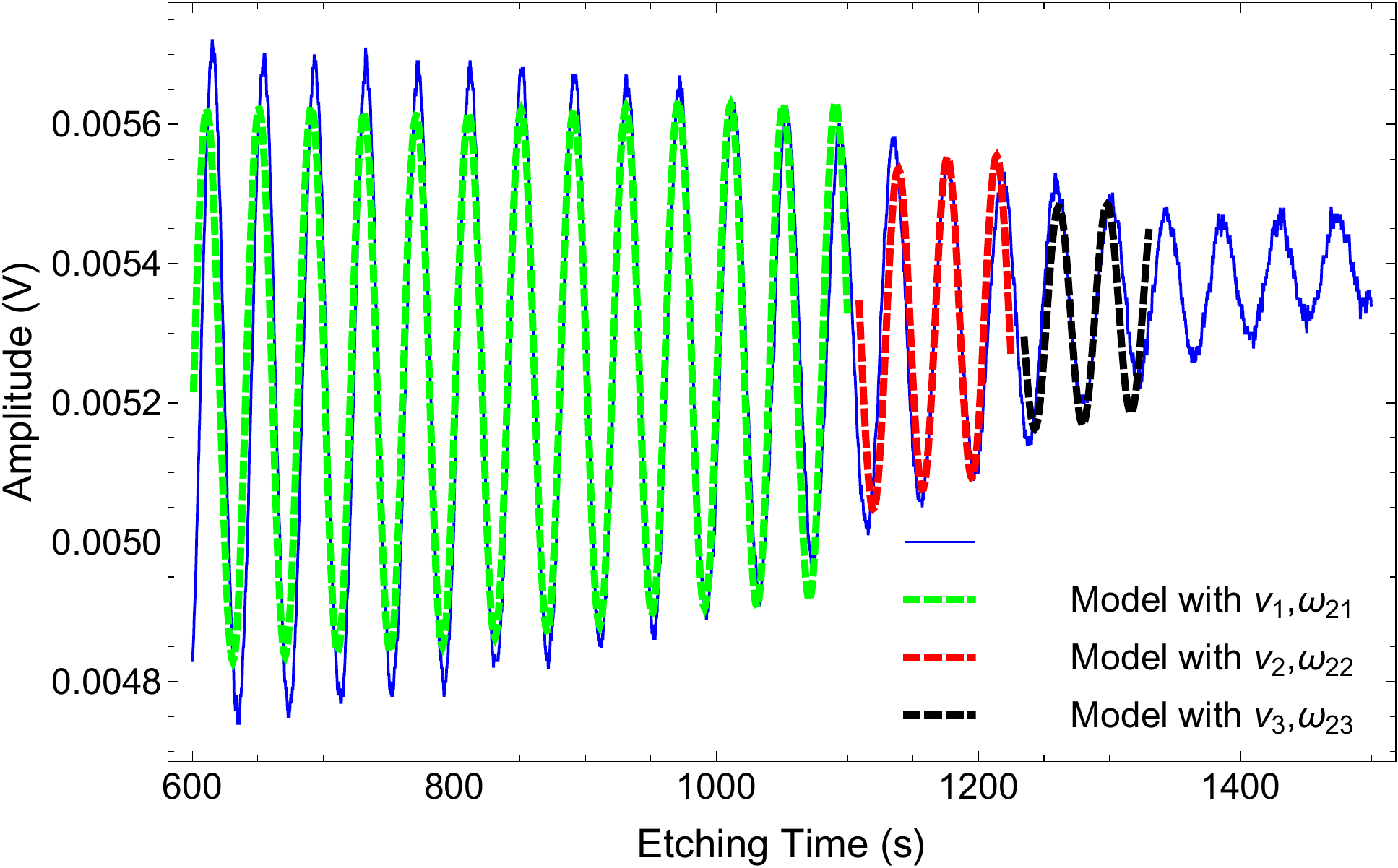}
\caption{(Color line) photoacoustic signal of experimental data (blue) and phenomenological model, for different values of $\omega_2$ and $v$. The fit has been done with the values $\omega_{21}=\pi/20\approx 0.157$ rad/s, $v_1=8$ nm/s (green), $\omega_{22}=0.167$ rad/s, $v_2=44$ nm/s (red), and $\omega_{23}= 0.170$ rad/s, $v_3=60$ nm/s (black).\label{P1v2}}}
\end{figure}

\begin{figure}[h]
{\centering
\includegraphics[scale=0.43]{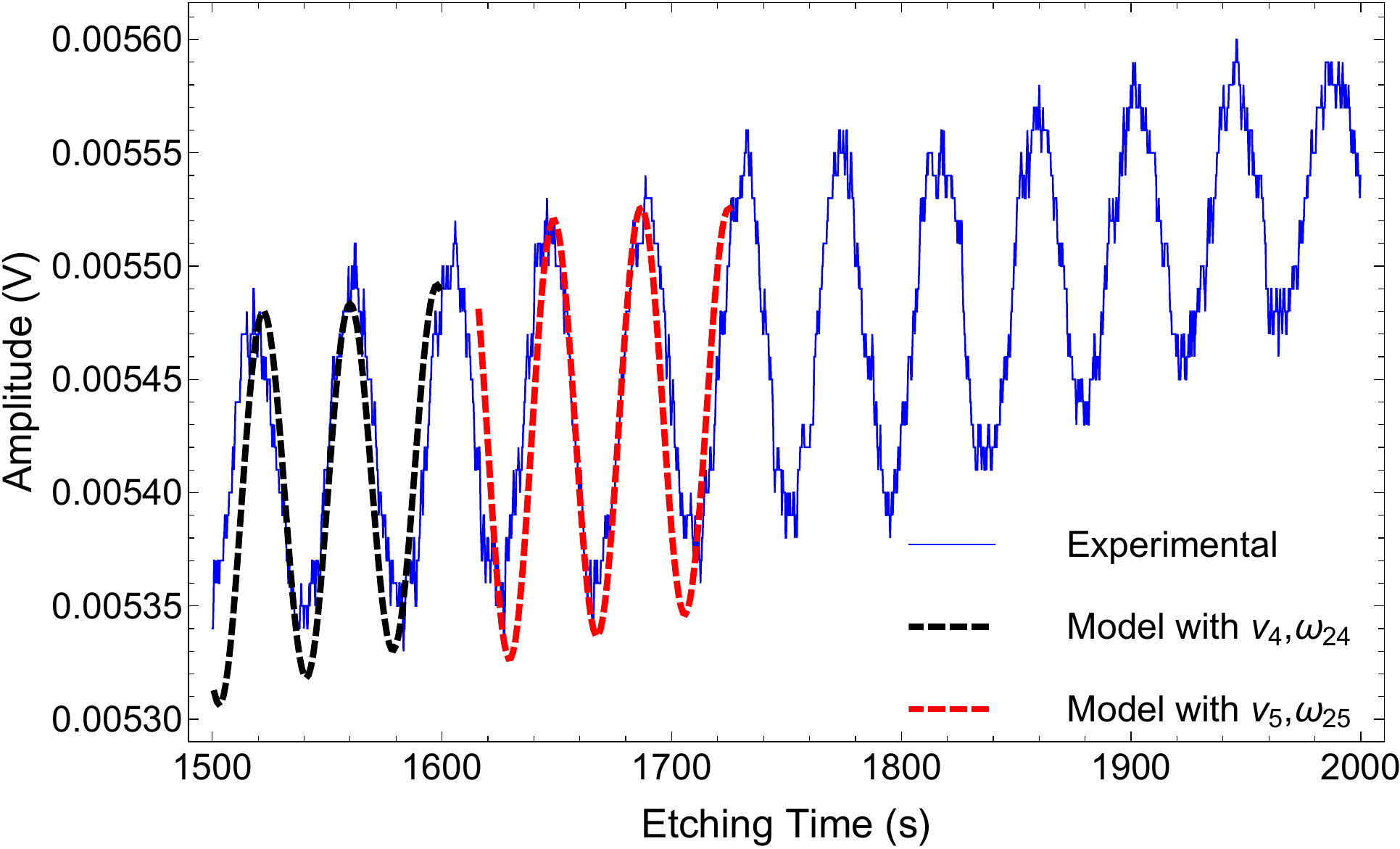}
\caption{(Color line) photoacoustic signal of experimental data (blue) and phenomenological model, for different values of $\omega_2$ and $v$. The fit has been done with the values $\omega_{24}=0.166$ rad/s, $v_1=50$ nm/s (black), and $\omega_{25}=0.165$ rad/s, $v_2=43$ nm/s (red) .\label{P2v2}}}
\end{figure}

It is necessary to take into account that the etching process also contributes to the PA signal as well as the DC component of the process (thermal inertia) which can be obtained from the experimental data using different criteria, such as: the minimum, the maximum, or the average of the PA amplitude signal of the characteristic growing process as a function of the etching time. In this case, the average criterion was used. It is important to note that this average temperature is the contribution of the DC and the etching heating. The inset in Fig.~\ref{experimental} shows the average temperature for the etching process in which it is noticeable that the temperature of the PA chamber increases. The changes in the pressure can be written as follows:
\bea
\Delta P=\Delta P_{\text{AC}}+\Delta P_{\text{DC}}+ \text{Etch}.
\label{Ptotal}
\eea
Figs.~\ref{P1} and~\ref{P2} exhibit a comparison between the experimental data (blue line) for 64 PA cycles and the theoretical results for  the amplitude obtained using the Eq.~(\ref{ReP}) and the values from Tables ~\ref{tabla1} and ~\ref{tabla2} (red line), as well as the DC contribution coming from the chemical reaction (green line). It can be noticed that there are regions  where the theoretical prediction does not fit the experimental data. It is taken into account that the velocity of the PS for growing with four PA cycles was determined earlier\cite{RCR}, indicating that the velocity of the etching is constant almost for a few cycles. The shift in some intervals of etching time evidence changes in the velocity of the PS formation. This behaviour can be associated with the spatial distribution of the impurities\cite{Mariovalexp1} and the variations  in the crystalline quality of the Si wafer.\cite{JAPMario}.\\
Figs.~\ref{P1v2} and~\ref{P2v2} a more detailed fit of  the experimental data is shown. For the simulations, the plot was divided into several regions, each one with its own parameter of etching velocity $v_i$ and frequency $\omega_{2i}$ in order to test the model when these variables were changed. The values of the parameters used are shown in Table~\ref{tabla2}. It can be seen that  at the beginning of the etching process the velocity of the PS growing is constant. As time passes, the etching velocity increases until a certain point in which $v$ decreases. This behaviour can be attributed to the depletion of the electrolyte. The results of the simulations show that a change in the etching velocity implies a change in the frequency $\omega_2$, and proves that the model allows the determination of these parameters which are the most important  information of the PA signal at any time interval of interest.

%%%%%%%%%%%%%%%%%%%%%%%%%%%%%%%%%%%%%%%%%%%%%%%%%%%%%%%%%%%%%%%%%%%%
%%%%%%%%%%%%%%%%%%%%%%%%%%%%%%%%%%%%%%%%%%%%%%%%%%%%%%%%%%%%%%%%%%%%
%%%%%%%%%%%%%%%%%%%%%%%%%%%%%%%%%%%%%%%%%%%%%%%%%%%%%%%%%%%%%%%%%%%%
From a physical point of view, the model reproduces the phenomenology of experimental the PA signal and predicts that the changes in the reflectance have a  frequency characteristic $(\omega_2)$ that governs the PA amplitude in some interval of  the etching time. An envelope function decays like exponential in accordance with the Lambert-Beer law. It gives information about the PS thickness, and the increases of the DC level is related to thermal inertia (Electrochemical etch + Joule Effect). Also, the thermal properties of the structure determined the behaviour of this envelope (see Eq.~\ref{Thetas}). In this case, the different frequencies ($\omega_2$) for the simulation were taken from the experimental data, to appreciate the accuracy of the model. The theoretical results exhibit an excellent comparison match with the experiment data. The reproduction of these cycles using our model is the most relevant information, for there is a direct relationship between the cycles and the velocity of  the PS formation.

\section{Summary and conclusions}\label{conclusions}
In summary the photoacoustic signal of the porous silicon formation has been studied by solving a one dimensional heat equation with an extension of the RG model. This modification takes into account three thermal sources: laser, reflectance changes,  and electrochemical reaction (etching process and Joule effect). The model reproduces in detail the growing conditions of PS. The frequency of the reflectance changes governs the PA signal; therefore, using this model, it is possible to determine the velocity of the etching process. Finally, from the inset in Fig.~\ref{experimental} the electrochemical reaction is continuous, and the increment of the PA amplitude can be considered as an exothermic reaction.

\begin{acknowledgments}
C. F. Ramirez-Guerrez and J. D. Castaño-Yepes want to thank Consejo Nacional de Ciencia y Tecnología México, for the financial support of their Ph.D. studies. We appreciate the electronic technical support of  J. A. Ramirez-Gutierrez M.Sc., and thank Beatriz Millan-Malo Ph.D., for her assistance in the simulations. We acknowledge the English editing by Agustín Ruiz Esparza Y Ballesteros and Felipe Cortes Uribe. This work was supported by PAPIIT IN115113 UNAM-M\'exico.

\end{acknowledgments}


\begin{thebibliography}{X}
\bibitem{Uhlir}A. Uhlir, The Bell System Technical Journal \textbf{35},333-347 (1956). %1
\bibitem{Foll}H. F\"{o}ll, M. Christophersen, J. Carstensen, and G. Hasse, Mat. Sci. Eng. R, \textbf{39}(4), 93-141 (2002).%2
\bibitem{Zhang}X. G. Zhang, J. Electrochem. Soc. \textbf{151}, C69-C80 (2004).%3
\bibitem{Zhao}Y. P. Zhao, Y. J. Wu, H. N. Yang, G. C. Wang, and T. M. Lu, Appl. Phys. Lett. \textbf{69}, 221 (1996). %4
\bibitem{RAO}A. V. Rao, F. Ozanam, and J. N. Chazalviel, J. Electrochem. Soc. \textbf{138}, 153-159 (1991).%5
\bibitem{ISA} M. Isaiev, K. Voitenko, V. Doroshchuka, D. Andrusenkoa, A. Kuzmicha, A. Skryshevskiia, V. Lysenkob, and R. Burbeloa, Phys. Procedia \textbf{70}, 586-589 (2015). %6
\bibitem{Gaburro} Z. Gaburro, C. J. Oton, P. Bettotti, L. Dal Negro, G. Vijaya Prakash, M. Cazzanelli, and L. Pavesi, J. Electrochem. Soc. \textbf{150}(6) C381-C384 (2003). %7
\bibitem{Foss}S. E. Foss, P. Y. Y. Kan, and T. G. Finstad, J. Appl. Phys. \textbf{97}, 114909 (2005). %8
\bibitem{RCR}C. F. Ramirez-Gutierrez, J. D. Casta\~no-Yepes and M. E. Rodriguez-Garcia, J. Appl. Phys. \textbf{119}, 185103 (2016).%9
\bibitem{REVMario}M. E. Rodr\'iguez-Garc\'ia, R. Vel\'aquez-Hern\'andez, M. L. Mendoza-L\'opez, D. M. Hurtado-Casta\~neda, K. M. Brie\~no-Enr\'iquez, and J. J. P\'erez-Bueno, Rev. Sci. Instrum. \textbf{78}, 034904 (2007).%10
\bibitem{Calderon} A. Calder\'on, J. J. Alvarado-Gil, Yu. G. Gurevich, A. Cruz-Orea, I. Delgadillo, H. Vargas, and L. C. M. Miranda, Phys. Rev. Lett. \textbf{79}, 5022 (1997). %11
\bibitem{Gutierrez}A. Guti\'errez, J. Giraldo, R. Vel\'azquez-Hern\'andez, M. L. Mendoza-L\'opez, D. G. Espinosa-Arbel\'aez, A. del Real, and M. E. Rodriguez-Garc\'ia, Rev. Sci. Instrum. \textbf{81}, 013901 (2010). %12
\bibitem {Espinosa}D. G. Espinosa-Arbel\'aes, R. V. Vel\'asquez-Hern\'andez, J. Petricioli-Carranco, R. Quintero-Torres, and M. E. Rodr\'iguez-Garc\'ia, Phys. Status Solidi C. \textbf{8}, 6, 1856-1859 (2011). % 13
\bibitem{JAPMario} M. E. Rodr\'iguez, A. Mandelis, G. Pan, and J. A. Garc\'ia, J. Appl. Phys. \textbf{87}(11), 8113-8121 (2000). %14
\bibitem{Dramicanic} M. D. Drami\'canin, Z. D. Ristovski, P. M. Nikoli\'c, D. G. Vasiljevi\'c, and D. M. Todorovi\'c, Phys. Rev. B \textbf{51}(20), 14226-14232 (1995).%15
\bibitem{Gersho}A. Ronsencwaig and A. Gersho, J. Appl. Phys. \textbf{47}(64), 64-69 (1976). %16
\bibitem{Martin} M. A. Green and M. J. Keevers, Prog. Photovoltaics \textbf{3}, 189-192 (1995). %17
\bibitem{Chen}Z. Chen, T.-Y. Lee, and G. Bosman, Appl. Phys. Lett. \textbf{64}, 3446 (1994).%18
\bibitem{Siapkas} C. C. Katsidis and D. I. Siapkas, Appl. Opt. \textbf{41} (19), 3978-3987 (2002). %19
\bibitem{Amato} G. Amato, G. Benedetto, R. Spagnolo, and M. Turnaturi, Phys. Stat. Sol. (a) \textbf{114}, 519 (1989). %20
\bibitem{Salnick} A. Salnick, A Mandelis, F. Funak, and C Jean,  Appl. Phys. Lett. \textbf{71} (11), 1531 (1997). %21
\bibitem{Mariovalexp2} M. E. Rodr\'iguez, A. Mandelis, G. Pan, L. Nicolaides, J. A. Garc\'ia, and Y. Riopel, ?J. Electrochem. Soc. \textbf{147} (2), 687-698, (2000).%18
\bibitem{Pinto} A. Pinto, H. Vargas, N. F. Leite, and L. C. M. Miranda, Phys. Rev. B \textbf{41}, 9971-9979 (1990). %%19
\bibitem{Burzo} M. G. Burzo, P. L. Komarov, P. E. Raad, 5th Int. Conf. On Thermal And Mechanical Simulations and Experiments in Micro-electronics and Micro-Systems  EuroSiinE2004, 269-276 (2004). % 20
\bibitem{Hand} J. C. Dixon, The Shock Absorber Handbook Second Edition, John Wiley and Sons Appendix B, 375 (2007). %21
\bibitem{Mariovalexp1} M. E. Rodr\'iguez, J. A. Garc\'ia, A. Mandelis, C. Jean, and Y. Riopel,  Appl. Phys. Lett. \textbf{74}, 2429 (1999). %22

\end{thebibliography}
\end{document}